\documentclass[a4paper,11pt,hyper]{JHEP3}
\usepackage{amsmath,latexsym,amssymb}
\usepackage{graphicx}
\usepackage{psfrag}
\usepackage[latin1]{inputenc}
\usepackage{subfigure}
\usepackage{nicefrac}
\usepackage{bbm}
\usepackage[stable]{footmisc}

\usepackage{comment}

\newcommand{\eq}[1]{\begin{equation}#1\end{equation}}
\newcommand{\spl}[1]{\begin{split}#1\end{split}}
\newcommand{\al}[1]{\begin{align}#1\end{align}}
\newcommand{\subeq}[1]{\begin{subequations}#1\end{subequations}}

\newcommand{\arXividhepth}[1]{\href{http://arxiv.org/abs/#1}arXiv:{\tt #1} [hep-th]}
\newcommand{\arXividother}[2]{\href{http://arxiv.org/abs/#1}arXiv:{\tt #1} [#2]}

\newcommand{\vols}{V_s}

\def\d{\text{d}}

\def\slashchar#1{\setbox0=\hbox{$#1$}           
\dimen0=\wd0                                 
\setbox1=\hbox{/} \dimen1=\wd1               
\ifdim\dimen0>\dimen1                        
\rlap{\hbox to \dimen0{\hfil/\hfil}}      
#1                                        
\else                                        
\rlap{\hbox to \dimen1{\hfil$#1$\hfil}}   
/                                         
\fi}

\def\Re           {{\rm Re\hskip0.1em}}
\def\Im           {{\rm Im\hskip0.1em}}

\newcommand{\boxedeq}[1]{
\begin{equation}
\fbox{
\rule[0.7cm]{0pt}{0pt}
$#1$
\rule[-0.45cm]{0pt}{0pt}
}
\end{equation}
}

\title{On the Cosmology of Type IIA Compactifications on SU(3)-structure Manifolds}

\author{Claudio Caviezel${}^{\diamondsuit}$, Paul Koerber${}^{\diamondsuit}$, Simon K\"ors${}^{\diamondsuit}$, Dieter L\"{u}st${}^{\diamondsuit\clubsuit}$,
 Timm Wrase$^{\diamondsuit}$ and Marco Zagermann${}^{\diamondsuit}$ \\

\begin{itemize}
  
\item  Max-Planck-Institut f\"ur Physik\\
F\"ohringer Ring 6, 80805 M\"unchen, Germany
  
\item  Arnold-Sommerfeld-Center for Theoretical Physics\\
Ludwig-Maximilians-Universit\"at M\"unchen\\
Theresienstra\ss e 37, 80333 M\"unchen, Germany
  \end{itemize}

\bigskip
E-mail: \email{caviezel,koerber,koers,luest,wrase,zagerman@mppmu.mpg.de \& dieter.luest@lmu.de }}

\abstract{We study cosmological properties of type IIA compactifications on orientifolds of SU(3)-structure manifolds with non-vanishing geometric flux. These compactifications give rise to effective 4D $\mathcal{N}=1$ supergravity theories that do not fall under some recently-proven no-go theorems against de Sitter vacua and slow-roll inflation. Focusing on a well-understood class of models based on coset spaces, however, we can use a refined no-go theorem that rules out de Sitter vacua and slow-roll inflation in all but one case. The refined no-go theorem uses the dilaton and a specific linear combination of the K\"ahler moduli, which is
different from the overall volume modulus. It puts a lower bound on the first slow-roll parameter: $\epsilon\geq2$.  The only case not ruled out is the manifold SU(2)$\times$SU(2), for which we indeed find critical points with $\epsilon$ numerically zero. However, all the points we could find have a tachyon corresponding to an eta-parameter $\eta\lesssim -2.4$. 
}


\preprint{MPP-2008-165 \\ LMU-ASC 64/08}

\begin{document}
\setcounter{footnote}{0}
\renewcommand{\thefootnote}{\arabic{footnote}}
\setcounter{section}{0}

\section{Introduction}
Several beautiful astrophysical measurements over the recent years have provided a fascinating and coherent
picture about the evolution and large scale structure of our universe. In particular, we now know that the universe
is spatially flat, $|\Omega-1| \ll 1$, and the latest CMB data from WMAP5 agree with an
almost scale-invariant spectrum with scalar spectral index $n_s=0.96\pm 0.013$ \cite{Komatsu:2008hk}. These data can be nicely explained by an epoch of cosmic inflation in the early universe \cite{Inflation,books}, modelled by a suitable effective scalar field theory for an inflaton field $\phi$, whose positive scalar potential $V(\phi)$ drives the nearly exponential expansion.
As the details of the CMB spectrum depend sensitively on the precise shape of the inflaton potential, it is an exciting possibility to try to use CMB measurements
as a powerful, and possibly unique, probe of the fundamental theory of matter and gravity in the early universe. In order to come closer to this goal,
one would like to constrain the huge space of possible effective field theories to a much smaller, more manageable set. This can be done  either by improving the
 available data sets, or by attempts to  consistently embed models of inflation in
fundamental theories of quantum gravity such as string theory. That the latter is in general nontrivial follows from, e.g., the
generic sensitivity of inflation models to even Planck suppressed corrections to the inflaton potential, which cannot be chosen at will
in a given UV-complete theory, but take on definite values. This sensitivity concerns in particular the flatness parameters
\eq{\spl{
& \epsilon \equiv \frac{M_P^2 \, g^{AB}\partial_{A}V \partial_{B}V}{2 \, V^2} \, , \\
& \eta \equiv \text{min eigenvalue} \left(\frac{M_P^2 \, \nabla^A \partial_B V}{V} \right) \, ,
}}
where we have displayed the general expressions for several scalar fields with $g^{AB}$ being the inverse scalar field metric.
The self-consistency of the standard slow-roll approximation (and relations such as
$n_s=1-6\epsilon+2\eta \,$) requires these parameters to be small,
\eq{
\epsilon \ll 1 \, , \qquad |\eta| \ll 1 \, .
}

Another important cosmological observation of the past decade is that also the present universe is in a state of accelerated expansion \cite{Supernova}, apparently driven by a non-vanishing
vacuum energy with an equation of state very close to that of a small and positive cosmological constant $\Lambda$. In an effective field theory setup, an asymptotic
de Sitter (dS) phase induced by a constant vacuum energy would correspond to a local minimum of the potential with
 $\epsilon = 0$ and $\eta >0$ at $V>0$.

The moduli of string compactifications are often considered as natural inflaton candidates, and various
inflation models have been proposed based on this idea (for recent reviews on the subject, see, e.g., \cite{Infreviews} and references therein). These models can roughly be divided into closed string inflation models, in which the inflaton is
a closed string modulus, and  open string (or brane-) inflation models, where the role of the inflaton is played
by a scalar describing some  relative brane distance or orientation.\footnote{Mixtures of open and closed string moduli have also been considered as inflaton candidates, e.g., in some variations of D3/D7-brane inflation \cite{D3D7}.}
In any such model, it is important to stabilize the orthogonal moduli, which one does not want to participate in inflation,
in particular if these orthogonal moduli correspond to potentially steep directions of the scalar potential.

In contrast to type IIB string theory, where cosmological model building and moduli stabilization are already quite advanced subjects (following the work of \cite{Kachru:2003aw,Kachru:2003sx,Giddings:2001yu}), comparatively little is known in type IIA string theory, even though type IIA models
 are very interesting for several good reasons:

First of all, type IIA orientifolds with  intersecting D6-branes (see e.g.~\cite{Blumenhagen:2006ci,modelbuildingreviews} for reviews and many more references)
offer good prospects for deriving the Standard Model from strings, as was recently
demonstrated in \cite{Gmeiner:2008xq}.\footnote{On the other hand, for recent progress in GUT model building in type IIB orientifolds see
\cite{Blumenhagen:2008zz}. Furthermore, there has recently been a lot of activity in model building in F-theory following the work of \cite{vafaF}.}
So, if cosmological aspects can likewise be modelled, one may study questions such as, e.g., reheating much more explicitly.

Second, in type IIA compactifications, all geometrical moduli can be already stabilized at the classical level by fluxes, albeit to AdS vacua in
four dimensions \cite{DeWolfe:2005uu,dkpz,fontaxions,vill}. The advantage of such models is their explicitness and the possibility to stabilize the moduli in a well-controlled regime
(corresponding to large volume and small string coupling) with power law parametric control (instead of logarithmic as in type IIB constructions along the lines of \cite{Kachru:2003aw}).


The main problem of such  type IIA compactifications is that there exist
already quite strong  no-go theorems against dS vacua and slow-roll inflation:
extending the earlier work \cite{hertzbergreview}, the authors of \cite{Hertzberg:2007wc} prove a no-go theorem against
small $\epsilon$ in type IIA compactifications on Calabi-Yau manifolds
with standard RR and NSNS-fluxes, D6-branes and O6-planes at large volume and with small string coupling. This no-go theorem uses the particular functional dependence of the corresponding scalar potentials on the volume modulus $\rho$ and the 4D dilaton $\tau$. More precisely, using the ($\rho$, $\tau$)-dependence, they show that the slow-roll parameter $\epsilon$ is at least $\frac{27}{13}$ whenever the potential is positive, ruling out slow-roll inflation in a near-dS regime, as well as meta-stable dS vacua.
As was already emphasized in  \cite{Hertzberg:2007wc}, however, the inclusion of other ingredients such as NS5-branes,
geometric fluxes and/or non-geometric fluxes evade the assumptions that underly this no-go theorem.
From these ingredients, especially geometric fluxes are quite natural in a type IIA context, as it is known
that D6-branes and fluxes in that case have a stronger backreaction than the ISD three-form fluxes in IIB \cite{Giddings:2001yu},
deforming the internal geometry away from a Calabi-Yau manifold. Following the mathematical work of \cite{torsionclasses} there is now
also a better understanding of the resulting SU(3)-structure manifolds and their application to compactifications with
fluxes, see e.g.\ \cite{SU3,gl,Grimm:2004ua,effectivepapers,minkpreduction,Ihl:2007ah,Robbins:2007yv,Ihl:2006pp,granareview}.
In \cite{Silverstein:2007ac}, a combination of geometric fluxes, KK5-branes and discrete Wilson lines was indeed argued to
allow for dS minima. These ingredients were used in \cite{Silverstein:2008sg}
 to construct  large field inflationary models with very
interesting experimental predictions.


In the recent work \cite{Haque:2008jz}, $F_{0}$ flux (i.e.\ non-vanishing Romans mass) and geometric flux
were identified as ``minimal'' additional ingredients in order to circumvent the no-go theorem of \cite{Hertzberg:2007wc}.
In the present paper,
we  discuss the question to what extent the recently constructed type IIA
$\mathcal{N}=1$ AdS$_4$ vacua with SU(3)-structure \cite{behr,lt,paltihouse,Aldazabal:2007sn,tomasiellocosets,Koerber:2008rx,adseffective}
can be used for inflation or dS vacua.
In particular, the coset models with SU(3)-structure
could  be candidates for circumventing the no-go theorem of \cite{Hertzberg:2007wc}, as they all have geometric fluxes and allow for non-vanishing Romans mass.
Specifically, we investigate whether the scalar
potentials in the closed string moduli sector that were already investigated in \cite{adseffective} can be flat enough in order to allow for inflation by one
of the closed string moduli. For this to be the case the parameter $\epsilon$ must be small enough in
some region of the positive closed string scalar potential. In addition, this analysis is also relevant for open string inflation
in these IIA vacua, since in this case we have to find closed string minima of the scalar potential,
i.e.\ $\epsilon=0$ somewhere in the closed string moduli space. Having a point with $\epsilon=0$ would also be a necessary condition for a
dS vacuum somewhere in moduli space.

For finding a small $\epsilon$, however, it is not relevant whether the effective field theory actually has a supersymmetric vacuum. Because of this we also extend our analysis to two more coset spaces that allow for an SU(3)-structure but do not admit a supersymmetric AdS vacuum.

The main result of our investigation is that we can apply, for all but one model, a refined no-go theorem of \cite{Flauger} that does \emph{not} just use the volume modulus  and the dilaton, but also some of the other K\"{a}hler moduli.\footnote{Problems with field directions orthogonal to the ($\rho$, $\tau$)-plane were also discussed in \cite{Haque:2008jz}, where attempts were made to construct dS vacua on manifolds that are products of certain three-manifolds.} These would not have been ruled out by the no-go theorem of \cite{Hertzberg:2007wc} (except for the example of positive curvature, which we already excluded in \cite{adseffective}).
Just as in \cite{Hertzberg:2007wc}, it is the epsilon parameter, i.e., first derivatives of the potential that cannot be made small.
Our results in particular show that it is important to make sure that the potential has a critical point (or small first derivative) in \emph{all} directions in moduli space. Moreover, the refined no-go theorem, just as the one of \cite{Hertzberg:2007wc}, is of a different nature than the no-go theorems developed in \cite{etanogo}, which assume a vanishing (or small) first derivative and then show that, under certain conditions, the eta parameter cannot be made small enough.

The coset model we do not rule out by a no-go theorem corresponds to the group manifold SU(2)$\times$SU(2). For this model, we can indeed show that points in moduli space with $\epsilon\approx  0$ exist. The points we could find, however, have a tachyonic direction corresponding to $\eta\lesssim-2.4$, making them useless for our intended phenomenological applications. We could not rule out the existence of analogous points with smaller $|\eta|$.

The organization of the paper is as follows: in section 2, we review the no-go theorems discussed in \cite{Hertzberg:2007wc} and \cite{Haque:2008jz}, paying particular attention to the role of geometric fluxes. In section 3, we discuss a refined no-go theorem of \cite{Flauger} that can be applied to SU(3)-structure compactifications with a certain form of the triple intersection  numbers. We also comment on the possibility of circumventing these no-go theorems by adding a number of additional ingredients. In section 4, we apply the no-go theorem of section 3 to the coset spaces and group manifolds with SU(3)-structure discussed in \cite{adseffective} and the two extra non-supersymmetric cosets, taken from \cite{Koerber:2008rx}. Our results are summarized in section \ref{Conclus}. Some technical details on the moduli space are relegated to appendix \ref{bubbles}.


\section{Geometric fluxes and no-go theorems in the volume-dilaton plane}
\label{Oldnogotheorem}
We start by reviewing previously derived no-go theorems  \cite{Hertzberg:2007wc} (see also \cite{Silverstein:2007ac,Haque:2008jz})
that exclude slow-roll inflation and dS vacua in the simplest compactifications of massive type IIA supergravity, focusing  in particular on the role played by the curvature of the internal space.

Classically, the four-dimensional scalar potentials of such compactifications may receive contributions from the NSNS $H_3$-flux, geometric fluxes\footnote{Geometric flux is not a terribly well-defined concept. For us the internal manifold will have geometric flux if the Ricci scalar $R$ is non-zero. In the special case of group and
coset manifolds, the geometric flux can be related to the structure constants $f^{\alpha}{}_{\beta\gamma}$.}, O6/D6-branes and the RR-fluxes\footnote{We use the democratic formalism for the RR-fluxes \cite{democratic}
introducing extra fluxes for $p=6,8,10$ and a duality constraint. Next, we impose the compactification ansatz $F^{\text{tot}}_p = F_p + \d \text{vol}_4 \wedge \tilde{F}_{p-4}$, where $F_p$ and $\tilde{F}_{p-4}$ have only internal indices. The duality condition allows then to express $\tilde{F}_{6-p}$ in terms of $F_p$ with $p=0,2,4,6$.} $F_p,\,p =0,2,4,6$  leading to, respectively, the following terms:
\begin{equation}\label{eq:pot}
V= V_3 + V_f + V_{O6/D6} + V_0 + V_2 + V_4 +V_6,
\end{equation}
where
$V_3, V_0, V_2, V_4, V_6 \geq 0$,
and
$V_f$ and $V_{O6/D6}$ can a priori have either sign.

In \cite{Hertzberg:2007wc} the authors studied the dependence of this scalar potential on the volume modulus $\rho =(\textrm{Vol})^{1/3}$ and the four-dimensional dilaton $\tau = e^{-\phi} \sqrt{\textrm{Vol}}$. Using only this $(\rho,\tau)$-dependence, they could derive a no-go theorem in the absence of metric fluxes that puts a lower bound on the first slow-roll parameter,
\begin{equation}
\epsilon \equiv \frac{K^{A\bar{B}}\partial_{A}V \partial_{\bar{B}}V}{V^2} \geq \frac{27}{13}, \quad \textrm{whenever }  V>0, \label{Bound1}
\end{equation}
where $K^{A\bar{B}}$ denotes the inverse K\"{a}hler metric, and the indices $A,B,\ldots$ run over all moduli fields.
This then not only excludes slow-roll inflation but also dS vacua (corresponding to $\epsilon=0$).

The lower bound (\ref{Bound1}) follows from the observation that
a linear combination of the derivatives with respect to $\rho$ and $\tau$ is always greater than a  certain positive multiple of the  scalar potential $V$.
More precisely, it is not difficult to obtain the general scaling behavior of these terms with respect to $\rho$ and $\tau$,
\eq{\label{scalings}
V_3\propto \rho^{-3}\tau^{-2}, \qquad V_{p}\propto \rho^{3-p}\tau^{-4}, \qquad V_{O6/D6}\propto \tau^{-3}, \qquad V_{f}\propto \rho^{-1}\tau^{-2} \, ,
}
which then implies for the scalar potential \eqref{eq:pot}
\eq{\label{nogoKachru}
  -\rho\frac{\partial V}{\partial \rho} -3\tau \frac{\partial V}{\partial \tau}=9V+\sum_{p={2,4,6}}pV_{p} -2V_{f}.
}
Hence, whenever the contribution from the metric fluxes $V_f$ is zero or negative, the right hand side in \eqref{nogoKachru} is at least equal to $9V$, which can then be translated to the above-mentioned lower bound $\epsilon \geq \frac{27}{13}$.
Avoiding this no-go theorem without introducing any new ingredients would thus require $V_f > 0$. Since $V_f \propto -R$, where $R$ denotes the internal scalar curvature, this is equivalent to demanding that the internal space has negative curvature. Since all terms in $V$ scale with a negative power of $\tau$ we see from \eqref{eq:pot} and \eqref{scalings} that we then also need $V_{O6/D6} <0$ to avoid a runaway.

Following a related argument in \cite{Haque:2008jz}, one can identify another combination of derivatives with respect to $\rho$ and $\tau$ that sets a bound for $\epsilon$:
\eq{\label{nogonew}
-3 \rho \frac{\partial V}{\partial \rho}-3 \tau \frac{\partial V}{\partial \tau} = 9 V +6 V_3 -6 V_0 +6 V_4+12 V_6 \ge 9 V -6 V_0.
}
In the case of vanishing mass parameter, we have $V_0=0$, and \eqref{nogonew} implies $\epsilon \geq \frac97$. Therefore, we need to have $V_f > 0,\, V_{O6/D6} <0$ and $V_0 \neq 0$ in order to avoid the above no-go theorems. Note that the only real restriction here is that we have to have a compact space with negative curvature since we are always free to turn on $F_0$-flux and to do an orientifold projection. Of the coset models discussed in \cite{adseffective} only the $\frac{\textrm{G}_{2}}{\textrm{SU}(3)}$ case has positive curvature over the whole parameter space and is thus ruled out by the above no-go theorem.
The other coset spaces studied in \cite{adseffective}, on the other hand,
do allow for negative curvature and are thus not affected by the no-go theorem of \cite{Hertzberg:2007wc}. This was already noted in \cite{adseffective} and justifies a closer look at these models. As already mentioned in the introduction we will also discuss two more coset spaces not analyzed in \cite{adseffective}, which do not admit a supersymmetric AdS vacuum.



\section{SU(3)-structure manifolds and a different no-go theorem}
\label{inflation}
The no-go theorems described in the previous section are very general in the sense that they do not assume anything specific about the geometry of the internal manifold apart from the possible presence of geometric fluxes, $p$-form fluxes and O6/D6-brane sources. An obvious way around the no-go theorem
is to look at internal spaces with geometric fluxes leading to negative curvature. In this section we will use the technology of SU(3)-structures and discuss a refined no-go theorem \cite{Flauger} that holds for an important subclass of compactifications with geometric fluxes.

On manifolds with SU(3)-structure, the structure group of the tangent bundle can be reduced to SU(3), which implies the existence
of a non-vanishing, globally defined spinor field $\eta_+$.\footnote{There are more general ways one can
decompose the 10D spinor fields into 6D and 4D spinors such that one still has a 4D $\mathcal{N}=2$ (or after orientifolding $\mathcal{N}=1$) supergravity theory.
These lead to so-called SU(3)$\times$SU(3)-structure and the low-energy supergravity is discussed in \cite{granasup,grimmsup,effective,cassanipotential}.}
In type II compactifications, the effective 4D theory can then be described by an $\mathcal{N}=2$ supergravity Lagrangian, which, under suitable orientifold projections, turns into an effective $\mathcal{N}=1$ theory.
From bilinears of the spinor field one can form  a globally defined real two-form, $J$, and a complex decomposable three-form, $\Omega$, which generalize, respectively, the K\"{a}hler form and the holomorphic three-form of a Calabi-Yau manifold:\footnote{On Calabi-Yau manifolds, the spinor field $\eta_+$
is covariantly constant with respect to the Levi-Civita connection, implying $\d J=\d \Omega=0$. In our case however, we have generically $\d J\neq 0$ and $\d \Omega\neq0$, and the decomposition of these quantities in terms of SU(3)-representations defines the intrinsic torsion classes \cite{torsionclasses} (see also \cite{granareview} for a review). These torsion classes then lead to a non-vanishing curvature scalar and can thus be related to the geometric flux.}
\begin{equation}
J_{mn}=i\eta_{+}^{\dagger}\gamma_{mn}\eta_{+}, \qquad \Omega_{mnp}=\eta_{-}^{\dagger}\gamma_{mnp}\eta_{+},
\end{equation}
where $\eta_{-}$ is the complex conjugate of $\eta_+$, and $\gamma_{mn}$, etc.\ denote the weighted antisymmetrized products of (internal) gamma matrices. Fierz identities then imply
\al{
\label{compat}
\Omega\wedge J = & 0 \, , \\
\Omega\wedge \Omega^{\ast} = & \frac{4i}{3}J^{3}\neq 0,
}
and the 6D volume-form is
\begin{equation}
\d \textrm{vol}_{6}= \frac{1}{6}J^{3} = -\frac{i}{8}\Omega\wedge\Omega^{\ast} \label{volumerelation} \, .
\end{equation}

In our concrete models, we will introduce a configuration of O6-planes such that the resulting
low-energy theory in four dimensions is $\mathcal{N}=1$ supersymmetric. In order not to be projected out, $H,F_2$ and $F_6$ should be odd, and $F_0$ and $F_4$ should be even under each orientifold involution. Furthermore, the condition that the orientifold projection does not fully break the supersymmetry requires $J$ and $\Re \Omega$ to be odd, and $\Im \Omega$ to be even under each orientifold involution.  The holomorphic variables of the low-energy $\mathcal{N}=1$ theory sit then in the expansion of the complex combinations $J_c= J - i \delta B$ and $\Omega_c = e^{-\Phi}\textrm{Im}\Omega + i \delta C_3$ \cite{Grimm:2004ua}, where $\delta B$ and $\delta C_3$ are fluctuations around the background of the NSNS two-form and RR three-form potential, respectively, and $\Phi$ is the 10D dilaton.

Just as in Calabi-Yau compactifications, one tries to expand the complexified $J_c$ and $\Omega_c$ in a suitable basis of forms to yield an analogue
of K\"{a}hler moduli, $k^{i}$, and complex structure moduli $\tilde{u}^{I}= \tau^{-1}u^{I}$:\footnote{\label{complrescale}Note that there are $h^{3+}$ moduli $z^I$ whose real parts $\Re{z^I} = u^I$ depend on $\tau$ and the $(h^{3+} - 1)$ complex structure moduli. We can take out the $\tau$-dependence and define $u^I = \tau \tilde{u}^I$. So strictly speaking the $\tilde{u}^I, I=1,\ldots, h^{3+}$ are not the complex structure moduli but rather functions of the $(h^{3+}-1)$ complex structure moduli that are not all independent.}
\subeq{\label{expansion1}\al{
 J_c = & (k^{i} - i b^{i}) Y_{i}^{(2-)} \equiv t^i Y_{i}^{(2-)}\, , \qquad (i=1, \ldots, h^{2-})\, ,  \\
\Omega_c = & (u^{I} + i c^{I})Y_{I}^{(3+)} \equiv  z^I Y_{I}^{(3+)} \, ,\qquad (I = 1, \ldots, h^{3+})\, .
}}
Here, $Y_{i}^{(2-)}$ and $Y_{I}^{(3+)}$ are a set of expansion forms with suitable parity under the orientifold
involution, as indicated by the superscript $+/-$. In contrast to the Calabi-Yau case, they are in general not harmonic, since $J_c$ and $\Omega_c$ are not necessarily closed. The numbers $h^{2-}$ and $h^{3+}$ therefore do not count harmonic forms
and should not be confused with the Hodge numbers.  The easiest way to satisfy the compatibility condition \eqref{compat} is to impose it for all choices of the
moduli $t^i$ and $z^I$, which implies for our basis forms
\eq{
\label{compatexp}
Y_{i}^{(2-)} \wedge Y_{I}^{(3\pm)} = 0 \, ,
}
for all $i$ and $I$.

As the expansion forms are in general not harmonic, the corresponding 4D scalars $t^{i}$, $z^{I}$ are usually not massless. There is thus no clear separation into massive and massless modes as in conventional Calabi-Yau compactifications, and the identification of a suitable expansion basis is generally an important open problem. One approach is to derive constraints on the basis from the requirement of obtaining an effective supergravity Lagrangian with an appropriate amount of supersymmetry ($\mathcal{N}=2$, or $\mathcal{N}=1$ after orientifolding). Another would be to find a consistent truncation, and a third, perhaps most physical, would be to try to establish a hierarchy of masses and keep only the light fields. See \cite{gl,Grimm:2004ua,granasup,minkpreduction,effectivepapers} for various approaches to this problem.
The position adopted in \cite{adseffective} is that on group manifolds or coset spaces, at least, a natural expansion basis is provided by the left-invariant forms. In our concrete examples we will restrict our discussions to these cases and hence only consider SU(3)-structures that are likewise left-invariant.
As expansion forms we will then choose for $Y_i^{(2-)}$ the left-invariant odd two-forms and for $Y_I^{(3+)}$ the left-invariant even three-forms.
In the concrete models of section \ref{app1} there are either no odd left-invariant five forms or we will arrange for them to be projected out
by orientifolding, so that the extra condition \eqref{compatexp} is trivial.

\subsection{The scalar potential in SU(3)-structure compactifications}

The main idea of the no-go theorems discussed in section \ref{Oldnogotheorem} is to find a subset of the moduli along which the first derivatives of the scalar potential cannot be simultaneously made sufficiently small for slow-roll inflation. In section \ref{Oldnogotheorem}, the relevant subset of the moduli consists of the overall volume modulus $\rho$ and the 4D dilaton $\tau$.

We will now closely follow \cite{Flauger} and discuss another no-go theorem that is very similar in spirit, but
concerns a different two-dimensional slice in moduli space that no longer involves the overall volume $\rho$, but a different (albeit related) K\"{a}hler modulus. In order to write down the dependence of the scalar potential on the moduli, we first define the K\"ahler potential \cite{Grimm:2004ua}\footnote{The constant last term
makes $e^{K}$ dimensionless.}
\eq{
\label{kahlereinstein}
K  = K_k  + K_c+ 3 \ln(8 \kappa_{10}^2 M_P^2) \, ,
}
where $K_k$ and $K_c$ are the parts containing, respectively, the K\"ahler and complex structure/dilaton moduli.
They are given by
\subeq{\al{
K_k  = & - \ln \int_M \, \frac{4}{3} J^3 = - \ln \left(8 \, \text{Vol}\right) \, , \\
\label{kahlercomplstruc}
K_c  = & - 2 \ln \int_M \, 2 \, e^{-\Phi}\Im \Omega \wedge e^{-\Phi} \Re \Omega = - 4 \ln \tau + \tilde{K}_c \, ,
}}
where $\tilde{K}_c$ only depends on the complex structure moduli $\tilde{u}^I$ of footnote \ref{complrescale}.
Also in the second line, $e^{-\Phi} \Re \Omega$ should be considered as a function of $e^{-\Phi}\Im \Omega$.\footnote{Indeed, $e^{-\Phi} \Omega$
should be a decomposable form, which, according to \cite{hitchin}, implies that one can find $e^{-\Phi }\Re \Omega$ from $e^{-\Phi}\Im \Omega$.}
Furthermore, from the relation (\ref{volumerelation}) we find
\begin{equation}
\textrm{Vol}= \frac{e^{-K_k}}{8} = \frac{\kappa_{ijk} k^i k^j k^k}{6} \, ,
 \end{equation}
where $\kappa_{ijk}$ denotes the triple intersection number, given in terms of the  odd two-forms $Y_i^{(2-)}$ as
\begin{equation}
\kappa_{ijk} = \int_M Y_i^{(2-)} \wedge Y_j^{(2-)} \wedge Y_k^{(2-)} \, .
\end{equation}
The K\"ahler metric is given in the standard way in terms of the K\"ahler potential\footnote{We will use the indices $I,J,\ldots$
for both the real coordinates $u^I$ as well as the complex coordinates $z^I$. There should be no confusion as the field itself is always
indicated. In the definition of the K\"ahler metric the bar-notation emphasizes that the second derivatives is with respect to the
complex conjugate. Likewise for the indices $i,j,\ldots$.}
\eq{
K_{I\bar{J}} = \frac{\partial^2 K_k}{\partial z^I \partial{\bar{z}}^{\bar{J}}} \, , \qquad
K_{i\bar{\jmath}} = \frac{\partial^2 K_c}{\partial t^i \partial{\bar{t}}^{\bar{\jmath}}} \, .
}
It is also convenient to introduce a rescaled inverse K\"ahler metric for the complex structure sector $\widehat{K}^{I\bar{J}} = \tau^{-2} K^{I\bar{J}}$ so that $\widehat{K}^{I\bar{J}}$ is independent of $\tau$.

For SU(3)-structure manifolds, geometric flux describes the non-closure of the forms $J$ and $\Omega$, i.e.\ the non-vanishing intrinsic torsion, and, following our expansion  (\ref{expansion1}), we parameterize them in terms of matrices $r_{iI}$ defined by \cite{Ihl:2007ah}
\begin{equation}
\d Y^{(2-)}_i = r_{iI} Y^{(3-)I} \, .
\end{equation}
 In terms of the above quantities, one finds \cite{Grimm:2004ua,Flauger}\footnote{See also \cite{cassanipotential} for a derivation of the scalar potential in the
general SU(3)$\times$SU(3)-case.} the following contribution to the scalar potential coming from $H_3$, the metric flux, $O6/D6$-branes and the RR $p$-forms\footnote{Since the explicit models we will consider later have $h^{2+} = 0$, there are no gauge fields arising from $C_3$, and we need not consider D-term contributions (which might in general arise in the presence of metric fluxes \cite{Ihl:2007ah,Robbins:2007yv}).}
\allowdisplaybreaks
\eq{\spl{
\label{eq:potential}
V_3 =& \frac{\widehat{K}^{I\bar{J}} a_I a_J}{\tau^2 \, \textrm{Vol}} \, , \\
V_f =& \frac{1}{2\tau^2 \, \textrm{Vol}} \left( \widehat{K}^{I\bar{J}} r_{iI} r_{jJ} k^i k^j - 2 (u^I r_{iI} k^i)^2 -4 \textrm{Vol} (\kappa_{i} k^i)^{-1 \, jk} r_{jI} r_{kJ} u^I u^J \right) \, , \\
V_{O6/D6} =& - \frac{u^I b_I}{\tau^3} \, , \\
V_0 =& \frac{c_0 \textrm{Vol}}{\tau^4}\, , \\
V_2 =& \frac{c_2^i \, c_2^j}{\tau^4 \, \textrm{Vol}} \left( \kappa_{ikl} k^k k^l \, \kappa_{jmn} k^m k^n -4 \, \textrm{Vol} \,  \kappa_{ijn} k^n \right) = \frac{16 \,  c_2^i \, c_2^j \,   \textrm{Vol} \, K_{i\bar{\jmath}}}{\tau^4}  \, ,
\\
V_4 =& \frac{c_{4i} \, c_{4j}}{\tau^4 \, \textrm{Vol}} \left(2 k^i k^j - 2 \textrm{Vol} (\kappa_{k} k^k)^{-1 \, ij} \right) = \frac{c_{4i} \, c_{4j}  K^{i\bar{\jmath}}}{2 \,  \tau^4 \, \textrm{Vol}} \, , \\
V_6 =& \frac{c_6}{\tau^4 \, \textrm{Vol}} \, .
}}
Here, $(\kappa_{k} k^k)^{-1 \, ij}$ denotes the inverse matrix of $\kappa_{kij}k^{k}$, and  the coefficients $a_I, b_I, c_2^i, c_{4i}$ and $c_0, c_6 \geq 0$ depend on the fluxes, $O6/D6$-brane charges and the axion moduli \footnote{We refer the interested reader to \cite{Flauger} for the explicit form of these coefficients.}.
As mentioned  before, the only contributions to the scalar potential that are not necessarily positive are $V_f$ and $V_{O6/D6}$, and we need $V_f$ to be positive in order to evade the no-go theorem of \cite{Hertzberg:2007wc} and $V_{O6/D6}<0$  to avoid a runaway in the $\tau$ direction.

\subsection{Special intersection numbers}
The coset examples of SU(3)-structure manifolds discussed in \cite{tomasiellocosets,Koerber:2008rx,adseffective} have special intersection numbers that allow a  split of the index $i$ of the K\"{a}hler moduli into $\{0,a\}, a=1,\ldots, (h^{2-} -1)$, such that the only non-vanishing intersection numbers are
\begin{equation}\label{eq:intersection}
\kappa_{0ab} \equiv X_{ab} \, .
\end{equation}
We now introduce a variable similar to $\rho$ above by defining
\begin{equation}
k^a = \sigma \chi^a \, ,
\end{equation}
where $\sigma$ is the overall scale of $(h^{2-}-1)$ K\"ahler moduli and the $\chi^a$ are constrained by $X_{ab} \chi^{a} \chi^b =2$.
The volume can now simply be written as $\textrm{Vol} = k^0 \sigma^2$.
We can then write $V_2$ and $V_4$ in terms of $X_{ab}$ rather than $\kappa_{ijk}$ and spell out the explicit dependence on $k^0$ and $\sigma$:
\eq{\spl{\label{eq:potential2}
V_2 &= \frac{4}{\tau^4 k^0 \sigma^2} \left( c_2^0 c_2^0 \sigma^4 + c_2^a c_2^b (k^0)^2 \sigma^2 (X_{ac} \chi^c X_{bd} \chi^d -X_{ab})\right), \\
V_4 &= \frac{2}{\tau^4 k^0 \sigma^2} \left( c_{40} c_{40} (k^0)^2 + c_{4a} c_{4b} \sigma^2 (\chi^a \chi^b - X^{-1 ab}) \right).
}}
It follows from the positivity of the K\"ahler metric (cf.\ the corresponding terms in (\ref{eq:potential})) and the orthogonality of $k^{0}$ and $k^{a}$ that the two terms in $V_2$ and the two terms in  $V_4$ above are all \emph{separately} positive definite.

\subsection{No-go theorems in the $(\tau,\sigma)$-plane}
\label{subsec:Nogo2}
We will now restrict ourselves to the moduli $\tau$ and $\sigma$ and adapt a no-go theorem from \cite{Flauger} that applies to all but one of the  coset models to be discussed in section \ref{app1}. To this end, we first isolate the contributions from $\sigma$ and $\tau$ to the slow-roll parameter $\epsilon$. To identify the contribution from $\sigma$, we use
 the explicit form of the inverse K\"ahler metric  for the K\"{a}hler moduli \cite{Grimm:2004ua}
\begin{equation}
K^{i\bar{\jmath}} = 2 k^i k^j -4 \textrm{Vol} (\kappa_k k^k)^{-1 \, ij}.
\end{equation}
For our special intersection numbers \eqref{eq:intersection} we then find
\begin{equation}
\frac14 K^{i\bar{\jmath}} \frac{\partial V}{\partial k^i} \frac{\partial V}{\partial k^j}= \left( k^0 \frac{\partial V}{\partial k^0} \right)^2 + \frac12 \left( \sigma \frac{\partial V}{\partial \sigma} \right)^2 + \left(\frac{\chi^a \chi^b}{2} - X^{-1 ab}\right) \frac{\partial V}{\partial \chi^a} \frac{\partial V}{\partial \chi^b},
\end{equation}
where $\left(\frac{\chi^a \chi^b}{2} - X^{-1 ab}\right)$ projects to the tangent space of the hyperplane $\chi^{a}X_{ab}\chi^{b}=2$.

The slow-roll parameter $\epsilon$ is
\begin{equation}
\epsilon = \frac{K^{A \bar{B}} \partial_{\phi^A} V \partial_{\bar{\phi}^{\bar{B}}} V}{V^2} = \frac{K^{A \bar{B}} \left( \partial_{\Re{\phi^A}} V \partial_{\Re{\phi^B}} V+\partial_{\Im{\phi^A}} V \partial_{\Im{\phi^B}} V\right)}{4 V^2} \, ,
\end{equation}
so that we have (using the $\tau$-dependence of $K_c$ spelled out in \eqref{kahlercomplstruc})
\begin{equation}
\epsilon \geq \frac{1}{V^2} \left( \frac12 \left( \sigma \frac{\partial V}{\partial \sigma} \right)^2 +\frac14 \left( \tau \frac{\partial V}{\partial \tau}\right)^2 \right) \geq \frac{1}{18 V^2} \left( \sigma \frac{\partial V}{\partial \sigma} +2 \tau \frac{\partial V}{\partial \tau}\right)^2.
\end{equation}
Thus, if we can show that
\begin{equation}
D V \equiv \left( -\sigma \partial_\sigma -2 \tau \partial_\tau \right) V \geq 6V, \label{newnogo}
\end{equation}
we would have
\begin{equation}
\epsilon \geq 2 \, ,  \quad \text{whenever } V > 0 \, ,
\end{equation}
and slow-roll inflation and dS vacua are excluded.

From \eqref{eq:potential} and \eqref{eq:potential2} we obtain, for intersection numbers of special form \eqref{eq:intersection},
\eq{\spl{\label{dvall}
DV_3 = & 6 V_3 \, ,\\
DV_{O6} = & 6 V_{O6}\, ,\\
DV_0 = & 6 V_0\, ,\\
DV_2 = & 6 V_2 + \text{positive term}\, ,\\
DV_4 = & 8 V_4 + \text{positive term}\, ,\\
DV_6 = & 10 V_6\, ,
}}
so (\ref{newnogo}), and hence $\epsilon \geq 2$, would follow if also $DV_{f}\geq 6 V_{f}$.
In \cite{Flauger} it was shown that the extra condition $r_{aI}=0$ would ensure that $D V_f = 6 V_f$, implying the no-go theorem. In the coset examples to be discussed in the next section, however, one has $r_{aI}\neq 0$. Therefore, we will explicitly check for each case separately whether $D V_f \geq 6 V_f$ is satisfied or not.
In order to do so, it is convenient to write
\begin{equation}\label{defU}
V_f = \frac{1}{2 \tau^2 \textrm{Vol}} U \, ,
\end{equation}
 so that
 \begin{equation}\label{dvfx}
 D V_f = 6 V_f + \frac{1}{2 \tau^2 \textrm{Vol}} D U = 6 V_f + \frac{1}{2 \tau^2 \textrm{Vol}} (-\sigma \partial_\sigma) U,
 \end{equation}
 and the no-go theorem applies if we can show that
\boxedeq{\label{nogo}
-\sigma \partial_\sigma U = -k^a \partial_{k^a} U \geq 0 \, .
}
Furthermore, if the inequality (\ref{nogo}) is strictly valid, Minkowski vacua are ruled out as well. This can be seen as follows. Using (\ref{dvall}) and (\ref{dvfx}), we obtain
\eq{
D V = 6 V + 2 V_4 + 4 V_6 + \frac{1}{2 \tau^2 \textrm{Vol}} (-\sigma \partial_\sigma) U + \text{positive terms} \, ,
}
so that for a vacuum, $D V = 0$, we find with (\ref{nogo})
\eq{
\label{potineq}
V = -\frac{1}{6} \left( 2 V_4 + 4 V_6 + \frac{1}{2 \tau^2 \textrm{Vol}} (-\sigma \partial_\sigma) U + \text{positive terms} \right) \le 0 \, .
}
So, if the inequality \eqref{nogo} holds strictly, also \eqref{potineq} holds strictly as well,  and Minkowski vacua are ruled out.

Indeed, we checked in particular that the coset models of
section \ref{app1} do not allow for \emph{supersymmetric} Minkowski vacua with left-invariant SU(3)-structure.
Strangely enough, this includes the case SU(2)$\times$SU(2) for which eq.~\eqref{nogo} can be violated. This model may still allow for a non-supersymmetric Minkowski vacuum. Incidentally, we checked also that there are no \emph{supersymmetric} Minkowski vacua on any of the cosets of table \ref{cosetmodels} with static SU(2)-structure, which falls outside
the scope of the SU(3)-structure based no-go theorem of this section.

\subsection{A comment on extra ingredients}\label{ingredients}

Some ingredients that are not taken into account in the original no-go theorem of \cite{Hertzberg:2007wc}, nor in the no-go
theorems of section \ref{subsec:Nogo2} are KK-monopoles, NS5-branes, D4-branes and D8-branes. Some of these ingredients
were used in constructing simple dS-vacua in \cite{Silverstein:2007ac}. KK-monopoles would drastically change the topology
and geometry of the internal manifold so that their introduction makes it difficult to obtain a clear ten-dimensional picture, hence we will not discuss this possibility further in this paper.
NS5-branes, D4-branes and D8-branes would contribute through their respective currents
$j_{\text{NS5}}$, $j_{\text{D4}}$ and $j_{\text{D8}}$ as follows to the Bianchi identities
\eq{\spl{
\d H = - j_{\text{NS5}} \, , \\
\d F_4 + H \wedge F_2  = - j_{\text{D4}} \, , \\
\d F_0 = - j_{\text{D8}} \, .
}}
Since $H$ and $F_2$ should be odd, and $F_0$ and $F_4$ even under all the orientifold
involutions, we find that $j_{\text{NS5}}$ is an odd four-form, $j_{\text{D4}}$ an even
five-form and $j_{\text{D8}}$ an even one-form. In the approximation of left-invariant
SU(3)-structure to be used in the next section, one should also impose these brane-currents
to be left-invariant (making the branes itself smeared branes). For the concrete models of section \ref{app1}
there are no such currents $j_{\text{NS5}}$, $j_{\text{D4}}$ or $j_{\text{D8}}$ with the appropriate properties under
all orientifold involutions, implying that NS5-branes, D4- and D8-branes cannot be used in these models.

Let us briefly mention that an F-term uplifting along the lines of O'KKLT \cite{Kallosh:2006dv,Kallosh:2006fm} by combining the coset models with the quantum corrected O'Raifeartaigh model will not be a promising possibility either. The O'Raifeartaigh model is given by $W_{\text{O}}=-\mu^2 S$ and $K_{\text{O}}=S\bar{S}-\frac{(S\bar{S})^2}{\Lambda^2}$. The model has a dS minimum for $S=0$ where $V_{\text{O}}\approx\mu^4$. We combine the two models as follows (the subscript IIA refers to the previously discussed flux and brane contributions)
\eq{
W=W_{\text{IIA}}+W_{\text{O}}\, , \qquad  \qquad K=K_{\text{IIA}}+K_{\text{O}} \, .
}
In lowest order in $S$ the total potential is then given by
\eq{\label{oIIA}
V \approx V_{\text{IIA}}+ e^{K_{\text{IIA}}} V_{\text{O}} + \ldots \, .
}
Note that we can then include the contribution of $V_{\text{up}}=e^{K_{\text{IIA}}} V_{\text{O}}$ in the no-go theorems, because the uplift potential $V_{\text{up}}$ scales like $F_6$,
\eq{
V_{\text{up}} = \frac{A_{\text{up}}}{\tau^4 \, \text{Vol}} \, .
}
Since we assume a positive uplift potential, $V_{\text{up}}>0$, the fact that $V_{\text{up}}$ scales like $F_6$ tells us that adding this uplift potential does
not help in circumventing the no-go theorems of section \ref{Oldnogotheorem} or section \ref{subsec:Nogo2}.



\section{Application: coset models}\label{app1}
In the previous section, we described a no-go theorem that rules out dS vacua and slow-roll inflation for
type IIA compactifications on certain types of SU(3)-structure manifolds, namely those for which one coordinate in the triple intersection
number $\kappa_{ijk}$ can be separated as in eq.~\eqref{eq:intersection} and the geometric fluxes induce the relation (\ref{nogo}).
While these seem to be quite strong assumptions, it turns out that a large part of the explicitly known examples
of non-trivial SU(3)-structure compactifications actually do fall into this category, as we will show in this section.

As a starting point we could consider internal manifolds, for which an explicit SU(3)-structure
compactification to a supersymmetric AdS space-time is known \cite{adseffective}. We are not directly  interested in this AdS vacuum,
but the moduli space of such a compactification might still
have regions where the scalar potential is positive and allows for local dS minima or suitable inflationary trajectories.
The explicitly known models with supersymmetric AdS vacua can be divided into two classes:
\begin{itemize}
\item Nilmanifolds (or ``twisted tori'')\footnote{Nilmanifolds are group manifolds of nilpotent groups, quotiented by a discrete group to make them compact. In the physics literature they are known as twisted tori, since they can be described as torus bundles
on tori. See \cite{scan} for a short introduction.}
\item Group manifolds and coset spaces based on semi-simple and U(1)-groups
\end{itemize}
In each case, only left-invariant SU(3)-structures are considered. The scalar curvature of a nilmanifold reads $R=-\frac{1}{4} f^{\gamma_1}{}_{\alpha_1\beta_1} f^{\gamma_2}{}_{\alpha_2\beta_2} g_{\gamma_1\gamma_2}g^{\alpha_1\alpha_2}g^{\beta_1\beta_2}$ in terms of the internal metric $g_{\alpha\beta}$ and the structure constants $f^{\gamma}{}_{\alpha\beta}$ of the nilpotent group. Apart from the torus, this is always negative so that, as discussed in section \ref{Oldnogotheorem}, the nilmanifolds provide prime candidates for avoiding the no-go theorem of \cite{Hertzberg:2007wc}. However, the only known nilmanifold example, next to the torus, allowing for an $\mathcal{N}=1$ AdS$_4$ solution is the Iwasawa manifold, which turns out to be T-dual to the torus solution \cite{kounnastsimpis,adseffective}. Having no geometric fluxes, the torus is ruled out by the no-go theorem of \cite{Hertzberg:2007wc}, and one expects this to be true then also for the Iwasawa manifold because of T-duality.

The second class of explicitly known examples, i.e.\ the group manifolds and coset spaces based on semi-simple
and U(1)-groups, will in the following simply be referred to as ``the coset models''. For reviews on coset spaces see \cite{cosetrev0,cosetrev1}. As was explained in \cite{Koerber:2008rx}, in order for a coset space $G/H$ to allow for an SU(3)-structure, the group $H$ should be contained in SU(3).\footnote{These coset spaces were already considered in the construction of heterotic string compactifications by \cite{Lust:1986ix}.} The list of such six-dimensional cosets and the corresponding structure constants were given in \cite{Koerber:2008rx} and are summarized in table \ref{cosetmodels}. Out of these only five lead to $\mathcal{N}=1$ AdS$_4$ solutions \cite{Koerber:2008rx}, as we have indicated in the table. In \cite{adseffective}, the low-energy effective actions for these compactifications were calculated, so we can now check whether the no-go theorems described above can be applied.
\begin{table}[ht]
\begin{center}
\begin{tabular}{|c|c|c|}
\hline
$G$ & $H$ & $\mathcal{N}=1$ AdS$_4$ \\
\hline
\hline
G$_2$ & SU(3) & Yes\\
\hline
SU(3)$\times$SU(2)$^2$ & SU(3) & No \\
\hline
\hline
Sp(2) & S(U(2)$\times$U(1)) & Yes \\
\hline
SU(3)$\times$U(1)$^2$ & S(U(2)$\times$U(1)) & No \\
\hline
SU(2)$^3\times$U(1) & S(U(2)$\times$U(1)) & No \\
\hline
\hline
SU(3) & U(1)$\times$U(1) & Yes \\
\hline
SU(2)$^2\times$U(1)$^2$ & U(1)$\times$U(1) & No \\
\hline
\hline
SU(3)$\times$U(1) & SU(2) & Yes \\
\hline
SU(2)$^3$ & SU(2) & No \\
\hline
\hline
SU(2)$^2\times$U(1) & U(1) & No \\
\hline
\hline
SU(2)$^2$ & 1 & Yes \\
\hline
\end{tabular}
\caption{\label{cosetmodels} All six-dimensional manifolds of the type $M=G/H$, where $H$ is a subgroup of SU(3) and $G$ and $H$ are both products of semi-simple and U(1)-groups. To be precise this list should be completed with the cosets obtained by replacing any number
of SU(2) factors in $G$ by U(1)$^3$.}
\end{center}
\end{table}

For the  cosmological applications we have in mind, however, it is not really relevant whether the effective field theories actually
have supersymmetric vacua. All we are really interested in here are regions in moduli space with positive potential energy, where supersymmetry is spontaneously broken anyway. It is thus interesting to consider also compactifications that do not allow for supersymmetric AdS vacua.
Still restricting to left-invariant SU(3)-structure there are only two more coset spaces of table \ref{cosetmodels}.
They are $\frac{\text{SU(2)}^2}{\text{U(1)}}\times \text{U(1)}$ and SU(2)$\times$U(1)$^3$.

In this section we will study each of these coset models separately, i.e.\ the seven models that allow for a left-invariant SU(3)-structure (including the five that also allow for a supersymmetric AdS vacuum). For the first four models the condition of left-invariance is very strong, and leaves only a very limited set of two-forms and three-forms as expansion forms, while there are no left-invariant one-forms nor five-forms. In these cases, we are able to show the no-go theorem without assuming that we introduce orientifolds. In the models with $G= \frac{\text{SU(2)}^2}{\text{U(1)}}\times \text{U(1)}$, SU(2)$\times$U(1)$^3$
and SU(2)$\times$SU(2) there are left-invariant one-forms and five-forms, which complicates matters. For instance, the condition \eqref{compatexp} becomes non-trivial. So in each of these cases, we will introduce enough orientifolds to eliminate one- and five-forms. Furthermore, we will make the simplification that the orientifolds are perpendicular to the coordinate frame, except for SU(2)$\times$SU(2), which does not allow for perpendicular orientifolds. It turns out that in that case one can choose the same orientifolds as in the supersymmetric AdS vacua of \cite{adseffective} leading
to the same expansion forms.

As we will show, dS vacua (as well as Minkowski) and slow-roll inflation are excluded for all these coset cases by the no-go theorem (\ref{nogo}), except for the case SU(2)$\times$SU(2).

Not requiring the presence of a supersymmetric SU(3)-structure AdS vacuum, one can consider, next to the nilmanifolds, also the solvmanifolds
i.e.\ the group manifolds of a solvable Lie group (for related work see \cite{Haque:2008jz}). Without the additional conditions on the left invariance as for cosets, 
both nilmanifolds and solvmanifolds will however also allow for one- and five-forms, and a large amount of fields. 
A detailed study of all cases would probably also require considering different choices of orientifolds on each manifold, and we leave this for future work.

\subsection{Models for which the no-go theorems hold}

\subsubsection{$\frac{\text{G}_2}{\text{SU(3)}}$}
For this case, one finds for the function $U$ of \eqref{defU}:
\begin{equation}
 U  \propto  -(k^1)^2 \, ,\label{vfg2}
\end{equation}
which is manifestly negative. This implies that $V_{f}$ itself is manifestly negative  so that the no-go theorem of \cite{Hertzberg:2007wc}, reviewed in section \ref{Oldnogotheorem}, already rules out this case \cite{adseffective}. All other coset models allow for values of the moduli such that $V_f>0$ and therefore require a more careful analysis using the refined no-go theorem of section~\ref{subsec:Nogo2}.

\subsubsection{$\frac{\text{Sp(2)}}{\text{S}(\text{U(2)}\times \text{U(1)})}$}
For this case, one has
\begin{equation}
 U \propto (k^2)^2 - 4 (k^1)^2 - 12 k^1 k^2 \, ,\label{vfsp}
\end{equation}
and the only non-vanishing intersection number is $\kappa_{112}$ and permutations thereof, so that $k^{2}$ plays the role of $k^{0}$, and we have
\begin{equation}
D U = -k^1 \partial_{k^1} U \propto 8 (k^1)^2 + 12 k^1 k^2 > 0 \, ,
\end{equation}
so that with $k^i > 0$ (because of metric positivity) the inequality~\eqref{nogo} is strictly satisfied and this model is ruled out.

\subsubsection{$\frac{\text{SU(3)}}{\text{U(1)}\times \text{U(1)}}$}
For this coset space, we have
\begin{equation}
 U \propto (k^1)^2+ (k^2)^2 + (k^3)^2 - 6 k^1 k^2 - 6 k^2 k^3 - 6 k^1 k^3 \, ,\label{vfsu3}
\end{equation}
and the non-vanishing intersection numbers are of the type $\kappa_{123}$ so that we can choose any one of the three $k$'s as $k^0$. We will choose $k^{0}$ to be  the biggest and assume without loss of generality  that this is $k^1$, i.e.\ that $k^1 \geq k^2,k^3$. We then find that
\begin{equation}
D U =(-k^2 \partial_{k^2}-k^3 \partial_{k^3}) U \propto (6 k^1-2k^2) k^2 + (6 k^1-2k^3) k^3+ 12 k^2 k^3 > 0,
\end{equation}
so that with $k^i > 0$ (because of metric positivity) this coset space is also ruled out by the no-go theorem (\ref{nogo}).

\subsubsection{$\frac{\text{SU(3)}\times \text{U(1)}}{\text{SU(2)}}$}
For this model, the function $U$ depends on an extra positive constant $\lambda$ related to the choice of orientifolds.
The function $U$ turns out to be
\begin{equation}
 U \propto (k^2)^2 (u^2)^2 \lambda - 8 k^1 k^2 |u^1 u^2| (1+ \lambda^2) \, ,\label{vfsu3u1}
\end{equation}
and the non-vanishing intersection numbers are of the form $\kappa_{112}$. Thus $k^{2}$ plays the role of $k^{0}$, and we find that
\begin{equation}
D U = -k^1 \partial_{k^1} U \propto 8 k^1 k^2 |u^1 u^2| (1+ \lambda^2) > 0,
\end{equation}
so that with $k^i > 0$ (because of metric positivity) this case is also ruled out.

\subsubsection{$\frac{\text{SU(2)}^2}{\text{U(1)}}\times \text{U(1)}$}

It was shown in \cite{Koerber:2008rx} that if the U(1) factor in the denominator
does not sit completely in the SU(2)$^2$, the resulting coset is equivalent to SU(2)$\times$SU(2),
so we exclude this possibility here, as the above notation already suggests. The internal manifold is then in fact equivalent to
$T^{1,1}\times$U(1). We choose the structure constants as follows
(this is $a=1$, $b=0$ compared to \cite{Koerber:2008rx})
\eq{\spl{
f^{1}{}_{23}=f^{7}{}_{45}=1, \quad \text{cyclic}, \, \\
f^3{}_{45}=f^2{}_{17}=f^1{}_{72}=1.
}}
The possible orientifolds that are perpendicular to the coordinate frame and compatible with these structure constants
are along\footnote{To be precise e.g.\ 123 means for the orientifold involution $e^1 \rightarrow e^1$, $e^2 \rightarrow e^2$, $e^3 \rightarrow e^3$, $e^4 \rightarrow -e^4$, $e^5 \rightarrow -e^5$, $e^6 \rightarrow -e^6$.}
\eq{
123\, , \qquad \, 345, \qquad 256\, , \qquad 146\, , \qquad 246\, , \qquad 156 \, .
}
In order to remove one-forms and five-forms, it turns out that we have to introduce two orientifolds, in particular one
of $\{123,345\}$ and one of $\{256,146,246,156\}$. It does not matter for the analysis which particular choice is made, but
for definiteness let us choose 345 and 256. We arrive then at the following expansion forms
\eq{\spl{
 \text{odd 2-forms:  } & \quad (e^{15}+ e^{24})\, ,\quad e^{36}\, , \\
 \text{even 3-forms: } & \quad e^{123}\, ,\quad (e^{256}-e^{146})\, ,\quad e^{345}\, , \\
}}
for \eqref{expansion1}.

There is always a change of basis such that we can assume $k^i > 0$. The conditions for metric positivity then become
\eq{\label{posdefg}
u^1 u^2 > 0 \, , \qquad u^1 u^3 > 0 \, .
}
$U$ becomes
\eq{
U  \propto \frac{-4 k^1 k^2 u^2 (u^1 + u^3) + (k^2)^2 \left[(u^1)^2 + (u^3)^2\right]} {2 \sqrt{u^1 u^3} |u^2|}\, .
}
The non-vanishing intersection number is $\kappa_{112}$ so that $k^2$ plays the role of $k^0$, and we get for (\ref{nogo}):
\eq{
D U=-k^1\partial_{k^1} U \propto \frac{2 k^1 k^2 u^2 (u^1 + u^3)}{\sqrt{u^1 u^3} |u^2|} > 0 \, ,
}
which is positive using the conditions (\ref{posdefg}). Hence, this case is ruled out as well.

\subsubsection{SU(2)$\times$U(1)$^3$}

In this case there are ten possible orientifold planes perpendicular
to the coordinate frame and compatible with the structure constants. It turns out that in order to remove the one- and five-forms
we have to choose at least three mutually supersymmetric orientifolds and that it does not matter for the analysis which ones we choose.
For definiteness, let us take
\eq{
123\, , \qquad \, 356, \qquad 246\, .
}
With these orientifolds, we get the following expansion forms to be used in \eqref{expansion1}
\eq{\spl{
\text{odd 2-forms:  } & \quad e^{16}  \, , \quad e^{25} \, , \quad  e^{34} \, , \\
\text{even 3-forms: } & \quad e^{123} \, , \quad e^{356} \, , \quad e^{264} \, , \quad e^{145} \, .
}}
Again there is always a change of basis such that we can assume $k^i > 0$. The positivity of the metric demands that
\eq{
\label{posdefSU2U1p3}
u^1 u^2 > 0 \, ,\quad  u^1 u^3 > 0 \, ,\quad  u^1 u^4 > 0 \, .
}
For the quantity $U$ as defined in (\ref{defU}) we get
\eq{
U  \propto \frac{(k^1 u^4)^2 + (k^2 u^3)^2 +(k^3 u^2)^2 - 2 k^1 u^4 k^2 u^3 - 2 k^1 u^4 k^3 u^2- 2 k^2 u^3 k^3 u^2 }{2 \sqrt{
 u^1 u^2 u^3 u^4}}\, .
}
The non-vanishing intersection number is $\kappa_{123}$ so that each $k^i$ can play the role of $k^0$. Without loss of generality we can assume $k^1 u^4 \ge k^2 u^3 > 0$, $k^1 u^4 \ge k^3 u^2 > 0$ and choose $k^0$ to be $k^1$. Thus we then find
\eq{
DU = (-k^2\partial_{k^2} -k^3\partial_{k^3}) U  \propto \frac{-(k^2 u^3-k^3 u^2)^2+k^1 u^4(k^2 u^3+k^3 u^2)}{\sqrt{u^1 u^2 u^3 u^4}} > 0 \, ,
}
so that we can also rule out this model.

\subsection{$\text{SU(2)}\times \text{SU(2)}$}
\label{SU2SU2case}
Thus far, we have found that $\epsilon \geq 2$ for all other cases. For the remaining coset space SU(2)$\times$SU(2), one finds
\eq{\spl{
U \propto& \sum_{i=1}^3 (k^i)^2 \left(\sum_{I=1}^4 (u^I)^2 \right) - 4 k^2 k^3 (|u^1 u^2| + |u^3 u^4|) \\
  &  - 4 k^1 k^2 (|u^1 u^4| + |u^2 u^3|)- 4 k^1 k^3 \left(|u^1 u^3| + |u^2 u^4|\right) \, ,\label{vfsu2}
}}
and the non-vanishing intersection numbers are of the form $\kappa_{123}$ so that we could choose any one of the $k$'s as $k^0$. However, it is not possible to apply the no-go theorem. This can be easily seen if we take for example $u^1 \gg u^2, u^3 ,u^4$. Then we have schematically $U \propto \vec{k}^2 (u^1)^2$ and $D U \propto - k^a k^a (u^1)^2 <0$. In \cite{Flauger} further no-go theorems have been derived but none of those apply to this case either. Let us therefore study it in more detail.

\subsubsection{Small $\epsilon$ for $\text{SU(2)}\times\text{SU(2)}$}
We have argued above that the known no-go theorems cannot be used to rule out small $\epsilon$ for this compactification. Indeed we will see that $\epsilon \approx 0$ is possible and there are dS extrema.

The superpotential and K\"{a}hler potential of the effective $\mathcal{N}=1$ supergravity have been derived in various ways in \cite{granasup,grimmsup,effective} (based on earlier work of \cite{GVW,gl,Grimm:2004ua}). Here we summarize the main formul\ae{} which will be used in the following. The superpotential for SU(3)-structure
reads in the Einstein frame
\eq{
\label{suppot}
\mathcal{W}  = \frac{1}{4 \kappa_{10}^2} \int_M \langle e^{i(J-i\delta B)}, F - i \d_{H} \left(e^{-\Phi} \Im \Omega +i \delta C_3 \right) \rangle \, ,
}
where $\langle \phi_1 , \phi_2 \rangle = \phi_1 \wedge \lambda(\phi_2)|_{\text{top}}$ is the Mukai pairing. $\lambda$ is the operator reversing the indices
of a form and $|_{\text{top}}$ selects the part of top dimension six, as necessary to integrate over the internal manifold $M$.
The scalar potential is given in terms of the superpotential via
\eq{\label{eq:spotential}
V(\phi,\bar{\phi}) = M_p^{-2} e^{\mathcal{K}} \left( \mathcal{K}^{A \bar{B}} D_A \mathcal{W} D_{\bar{B}} \mathcal{W}^* - 3 |\mathcal{W}|^2 \right) \, .
}
In order to eliminate the one- and five-forms we must introduce at least three mutually supersymmetric orientifolds, compatible with the structure
constants. We can then always perform a basis transformation so that the odd two-forms and odd/even three-forms
are the same as in \cite{adseffective} and read\footnote{There are no even two-forms for our choice of orientifold involutions.}
\eq{\label{basisSU2SU2}\spl{
Y^{(2-)}_1 = & e^{14}, \qquad Y^{(2-)}_2 = e^{25}, \qquad Y^{(2-)}_3 = e^{36}, \\
Y^{(3-)1} = & \frac14 \left( e^{156} - e^{234} - e^{246} + e^{135} + e^{345} - e^{126} + e^{123} - e^{456}\right), \\
Y^{(3-)2} = & \frac14 \left( e^{156} - e^{234} + e^{246} - e^{135} - e^{345} + e^{126} + e^{123} - e^{456}\right), \\
Y^{(3-)3} = & \frac14 \left( e^{156} - e^{234} + e^{246} - e^{135} + e^{345} - e^{126} - e^{123} + e^{456}\right), \\
Y^{(3-)4} = & \frac14 \left( -e^{156} + e^{234} + e^{246} - e^{135} + e^{345} - e^{126} + e^{123} - e^{456}\right), \\
Y^{(3+)}_1 = & \frac12 \left( e^{156} + e^{234} - e^{246} - e^{135} + e^{345} + e^{126} + e^{123} + e^{456}\right), \\
Y^{(3+)}_2 = & \frac12 \left( e^{156} + e^{234} + e^{246} + e^{135} - e^{345} - e^{126} + e^{123} + e^{456}\right), \\
Y^{(3+)}_3 = & \frac12 \left( e^{156} + e^{234} + e^{246} + e^{135} + e^{345} + e^{126} - e^{123} - e^{456}\right), \\
Y^{(3+)}_4 = & \frac12 \left( -e^{156} - e^{234} + e^{246} + e^{135} + e^{345} + e^{126} + e^{123} + e^{456}\right),
}}
where the $e^\alpha$ $(\alpha=1,\ldots, 6)$  are a basis of left-invariant 1-forms, and we use the shorthand notation  $e^{14} = e^1 \wedge e^4$ etc. The $e^\alpha$ satisfy
\begin{equation}
\d e^\alpha = -\frac12 f^\alpha{}_{\beta\gamma} e^\beta \wedge e^\gamma\, ,
\end{equation}
where the structure constants for $\text{SU(2)}\times\text{SU(2)}$ are $f^1{}_{23} = f^4{}_{56} = 1$, cyclic\footnote{This model can be thought of as a twisted version of $T^6/(\mathbb{Z}_2 \times \mathbb{Z}_2)$ as discussed in \cite{Aldazabal:2007sn}. In that paper the authors focused on moduli stabilization and model building while we are interested in the cosmological aspects.}. From this we find
\begin{equation}\label{Matrixexplicit}
\d Y^{(2-)}_i = r_{iI} Y^{(3-)I}, \quad \text{with} \quad r = \left(
                                                               \begin{array}{cccc}
                                                                 1 & 1 & 1 & -1 \\
                                                                 1 & -1 & -1 & -1 \\
                                                                 1 & -1 & 1 & 1 \\
                                                               \end{array}
                                                             \right).
\end{equation}
In terms of the above expansion forms, we can again define the complex moduli as in
(\ref{expansion1}).
The positivity of the metric demands
\eq{
u^1 u^2 < 0 \, ,\quad  u^3 u^4 < 0 \, ,\quad  u^1 u^4 < 0 \, .
}
Next we turn to the choice of background fluxes. As explained in appendix \ref{bubbles}, for the part of the moduli space where $H$ is non-trivial
in cohomology, $p \neq 0$ (see below), the most general form of the background fluxes is
\allowdisplaybreaks
\subeq{\label{Fluxessu2su2}\al{
F_0  = &  m,\\
F_2  = & m^i Y^{(2-)}_i,\\
F_4  = & 0,\\
F_6  = & 0,\\
H  = & p \left( Y^{(3-)}_1 + Y^{(3-)}_2-Y^{(3-)}_3 + Y^{(3-)}_4 \right).
}}
Plugging in these background values for the fluxes together with the expansion \eqref{expansion1} in terms of the basis \eqref{basisSU2SU2},
we find for the superpotential \eqref{suppot}
\eq{
\mathcal{W}  = \vols(4\kappa_{10}^2)^{-1} \left(m^1 t^2 t^3 + m^2 t^1 t^3 + m^3 t^1 t^2 - i m t^1 t^2 t^3 - i p (z^1+z^2-z^3+z^4) + r_{iI} t^i z^I \right) \ ,
}
and the K\"ahler potential
\begin{equation}
K= - \ln{\prod_{i=1}^3 \left( t^i+\bar{t}^i \right)} - \ln{\prod_{I=1}^4 \left( z^I+\bar{z}^I \right)} + 3 \ln \left( \vols^{-1} \kappa_{10}^2 M_P^2 \right) + \ln 32 \, ,
\end{equation}
where $\vols=-\int_M e^{123456}$.
Note that the superpotential depends on all the moduli so there are no flat directions in this model.

It is straightforward to calculate the scalar potential \eqref{eq:spotential} and the slow-roll parameter $\epsilon$ from the K\"ahler and superpotential. Although we cannot analytically minimize $\epsilon$, we can do it numerically. One particular solution with numerically vanishing $\epsilon$ is
\allowdisplaybreaks
\eq{\spl{
& m^1 = m^2 = m^3 = L \, ,\qquad m = 2 \, L^{-1} \, , \qquad p = 3 \, L^2 \, , \\
& k^1 = k^2 = k^3 \approx .8974 \, L^2 \, , \qquad b^1 = b^2 = b^3 \approx - .8167 \, L^2 \, , \\
& u^1 \approx 2.496 \, L^3 , \qquad u^2 = - u^3 =u^4 \approx -.05667 \, L^3 \, , \\
& c^1\approx -2.574 \, L^3 \, , \qquad  c^2=-c^3=c^4 \approx .3935 \, L^3 \, ,
}}
where $L$ is an arbitrary length. While we can use $L$ to scale up our solution with respect to the string length $l_s$, we stress that this does not
correspond to a massless modulus, as it also changes the fluxes.

We conclude that in this case there is no lower bound for $\epsilon$. To obtain a trustworthy supergravity solution 
we would have to make sure that the internal space is large compared to the string length and that the string coupling is small (for which we could use our freedom in $L$). Furthermore, in the full string theory the fluxes have to be properly quantized. 
Although we do not think that this would prevent small $\epsilon$, we did not try to find such a solution because all the solutions with 
vanishing $\epsilon$ we found have a more serious problem, namely that $\eta \lesssim -2.4$. The eigenvalues 
of the mass matrix turn out to be generically all positive except for one, with the one tachyonic direction being a mixture of all the light fields, in particular the axions. This means that we have a saddle point rather than a dS minimum. A similar instability was found in related models in \cite{Flauger}. 

In \cite{etanogo}, a no-go theorem preventing dS vacua and slow-roll inflation was derived by studying the eigenvalues of the mass matrix. Allowing for an arbitrary tuning of the superpotential it was shown that for certain K\"ahler potentials the Goldstino mass is always negative. For the examples we found, this mass is always positive so that the no-go theorem of \cite{etanogo} does not apply. This means that allowing for an arbitrary superpotential it should be possible to remove the tachyonic direction.  In our
case, however, the superpotential is of course not arbitrary.

Since the no-go theorems against slow-roll inflation do not apply and we have found solutions with vanishing $\epsilon$, we checked whether our solutions allow for small $\eta$ in the vicinity of the dS extrema. Unfortunately, this is not the case. In fact, we found that $\eta$ does not change much in the vicinity of our solutions where $\epsilon$ is still small.

It would be very interesting to study the $\text{SU(2)}\times\text{SU(2)}$ model further to check whether one can prove that there is always at least one tachyonic direction or whether it allows for metastable dS vacua after all. Understanding this tachyonic direction better should also allow to decide whether there are points in the moduli space that allow for slow-roll inflation in this model.


\section{Conclusions and outlook}\label{Conclus}
Type IIA compactifications on orientifolds of SU(3)-structure manifolds with fluxes and D6-branes are phenomenologically interesting because they lead to effective 4D $\mathcal{N}=1$ supergravity actions with rich potentials for the moduli. These potentials have a dilaton-volume dependence that forbids dS vacua or slow-roll inflation unless the compact space has a negative scalar curvature induced by the geometric fluxes (or other more complex ingredients are introduced \cite{Hertzberg:2007wc,Silverstein:2007ac,Haque:2008jz}).

Motivated by this, we analysed a class of explicitly known SU(3)-structure compactifications with fluxes and O6/D6-sources for which the full scalar potential can be written down in closed form.
The manifolds we studied are those coset spaces or group manifolds based on  semi-simple and U(1) groups that admit a left-invariant SU(3)-structure \cite{Koerber:2008rx}.

As indicated in table \ref{cosetmodels}, five out of these seven manifolds allow for 4D $\mathcal{N}=1$ AdS solutions that solve the full 10D field equations of massive IIA supergravity \cite{Koerber:2008rx}.
Apart from a particular nilmanifold (the Iwasawa manifold) and tori, these are, to the authors best knowledge, the only explicitly known examples of this type. Using the 4D  effective action worked out in \cite{adseffective}, we could rule out dS (as well as Minkowski) vacua and slow-roll inflation elsewhere in moduli space for four of these coset spaces
by using a refined no-go theorem that probes the scalar potential also along a K\"{a}hler modulus different from the overall volume modulus (see also \cite{Flauger}). Just as the no-go theorem of \cite{Hertzberg:2007wc}, this no-go theorem works by establishing a certain lower bound on the first derivatives of the potential, and hence the epsilon parameter, for $V\geq 0$. It is thus different in spirit from the no-go theorems given in \cite{etanogo}, which assume a small first derivative and consider consequences for the second derivatives, i.e.\ the eta parameter.

 The only  coset space that allows for supersymmetric vacua and that is not directly ruled out by any known no-go theorem is then the group manifold SU(2)$\times$SU(2). For this case, we were indeed able to find critical points (corresponding to numerically vanishing $\epsilon$) with positive energy density, but only at the price of a tachyonic direction, corresponding to a large negative eta-parameter, $\eta\lesssim -2.4$. Interestingly, this
tachyonic direction does not correspond to the one used in the different types of no-go theorems of \cite{etanogo}.
 As our numerical search was not exhaustive, however, we cannot completely rule out the existence of dS vacua or inflating regions for this case.
Since this case also does not allow for a supersymmetric Minkowski vacuum as mentioned below (\ref{potineq}), our discussion covers all SU(3)-structure compactifications on semi-simple and U(1) cosets that have a supersymmetric vacuum.

Furthermore, we also studied the remaining two coset spaces of table \ref{cosetmodels} that do admit an SU(3)-structure but no supersymmetric AdS vacuum.
Choosing for simplicity the O-planes such that one-forms are projected out and restricting to O-planes perpendicular to the coordinate frame, we could again use
the refined no-go theorem of section \ref{subsec:Nogo2} to rule out dS vacua and slow-roll inflation for both of these cases as well.

Our results show that a negative scalar curvature and a non-vanishing $F_0$ is in general not enough to ensure dS vacua or inflation (as also noted in \cite{Haque:2008jz}), and we give a geometric criterion that allows one to separate interesting SU(3)-structure compactifications from non-realistic ones.

Our study could be extended in several directions. For one thing, it would be extremely interesting to find explicit SU(3)-structure manifolds that
do not fall under the class of coset spaces we have discussed here  and to investigate their usefulness for cosmological
applications along the lines of this paper. The most obvious class of manifolds to study systematically
would be the nil- and solvmanifolds. Another interesting direction might be the study of compactifications on manifolds with $\mathcal{N}=1$ spinor ans\"{a}tze more general than the SU(3)-structure case \cite{granaN1}. Concerning the SU(2)$\times$SU(2) model discussed in our paper,
one might try to either find a working dS minimum, or rule it out based on another no-go theorem, perhaps by using methods similar in spirit to \cite{etanogo}, although a direct application of their results to this case does not seem possible.
Following \cite{Silverstein:2007ac,Silverstein:2008sg} or \cite{Saueressig:2005es,Davidse:2005ef,Palti:2008mg}, one could also try to incorporate additional structures such as NS5-branes or quantum corrections of various types. In section \ref{ingredients}, however, we found that at least for our models, the following additional ingredients cannot be added or
do not work: NS5-, D4- and D8-branes as well as an F-term uplift along the lines of O'KKLT \cite{Kallosh:2006dv,Kallosh:2006fm}. Perhaps also methods similar to the ones in  \cite{Lust:2008zd} for non-supersymmetric  Minkowski or AdS vacua  might be useful for the direct 10D construction of dS compactifications. There is certainly a lot to improve about our understanding of cosmologically realistic compactifications of the type IIA string!

\begin{acknowledgments}
We would like to thank Davide Cassani, Thomas Grimm, Jan Louis, Luca Martucci, Erik Plauschinn, Dimitrios Tsimpis, Alexander Westphal, and in particular Raphael Flauger, Sonia Paban and Daniel Robbins for useful discussions and correspondence. This work is supported by the Transregional Collaborative Research Centre TR33 ``The Dark Universe'' and the Excellence Cluster ``The Origin and the Structure of the Universe'' in Munich. C.~C., P.~K., T.~W. and M.~Z.~are supported by the German Research Foundation (DFG) within the Emmy-Noether-Program (Grant number ZA 279/1-2).
\end{acknowledgments}

\begin{appendix}

\section{Labelling the disconnected bubbles of moduli space by flux quanta}
\label{bubbles}

In section \ref{SU2SU2case} we will search for a configuration with small $\epsilon$ somewhere in the moduli space. As we will
argue in a moment, this moduli space consists of different disconnected ``bubbles'', i.e.\ these bubbles are such that
it is not possible to reach another bubble by finite fluctuations of the moduli fields. The approach of \cite{adseffective}
of starting from a supersymmetric configuration and expanding around it is inadequate for studying the whole configuration space since
on the one hand, there will be bubbles that do not contain a supersymmetric configuration, while on the other hand, there are bubbles that
contain more than one supersymmetric configuration. In fact, in section \ref{SU2SU2case} we find configurations with $\epsilon \approx 0$ and $V>0$ in bubbles that do not
allow for supersymmetric AdS vacua. We follow here the standard approach of classifying the moduli space by flux quanta, which is however complicated by the presence of Romans mass, $H$-field and O6-plane source.

Classifying the different bubbles in terms of fluxes amounts to finding configurations that solve the Bianchi identities
\subeq{\label{biIIA}\al{
\d H & = 0 \, , \label{biIIA1} \\
\d F_0 & =  0 \, , \label{biIIA2} \\
\d F_2 + mH & = -j_3 \, , \label{biIIA3} \\
\d F_4 + H \wedge F_2 & = 0 \, , \label{biIIA4}
}}
while two configurations are considered equivalent if they are related by a fluctuation
of the moduli fields, which after imposing the orientifold projection (and assuming it removes one-forms) is given by \cite{adseffective}
\allowdisplaybreaks
\subeq{\label{IIAflucori}\al{
\delta H & = \d \delta B \, , \label{IIAflucori1} \\
\delta F_0 & = 0 \, ,  \label{IIAflucori2} \\
\delta F_2 & = - m \delta B \, , \label{IIAflucori3} \\
\delta F_4 & = \d \delta C_3 - \delta B \wedge (F_2 + \delta F_2) - \frac{1}{2} m (\delta B)^2 \, , \label{IIAflucori4} \\
\delta F_6 & = H \wedge \delta C_3 - \delta B \wedge (F_4 + \delta F_4) - \frac{1}{2} (\delta B)^2 \wedge (F_2 + \delta F_2) - \frac{1}{3!} m (\delta B)^3 \, . \label{IIAflucori5}
}}
In other words, we want to find representatives of the cohomology of the Bianchi identities \eqref{biIIA} modulo pure fluctuations of the potentials \eqref{IIAflucori}.

From eqs.\ (\ref{biIIA1}), \eqref{biIIA2}, \eqref{IIAflucori1} and \eqref{IIAflucori2} follows
immediately that $H \in H^3(M,\mathbb{R})$ and $F_0$ constant. To analyse \eqref{biIIA3} and \eqref{IIAflucori3} we take the point of
view that we choose the flux $F_2$, which then determines the source $j_3$. In fact, if $F_0 \neq 0$ the flux $F_2$ is only determined
up to a closed form, since the fluctuation $\delta B$ was from \eqref{IIAflucori1} also only determined up to a closed form, which can then be used in \eqref{IIAflucori3}.
Moving on to $F_4$, we find that in eq.~\eqref{biIIA4} $H \wedge F_2=0$, since we assumed there were
no even five-forms under all the orientifold involutions. Moreover, with the fluctuations $\delta C_3$ we can remove the exact part of $F_4$
so that $F_4 \in H^4(M,\mathbb{R})$. This however, leaves the closed part of $\delta C_3$ undetermined, which, if we have chosen $H$ non-trivial,
can in the SU(2)$\times$SU(2) case be used to put $F_6=0$.

Taking into account the parity requirements under the orientifold involution, we find for the case of SU(2)$\times$SU(2)
the general form of the background eq.~\eqref{Fluxessu2su2} when $H$ is non-trivial. If $H$ is trivial one must allow for non-zero $F_6$.

\end{appendix}

\end{document}